\documentstyle[preprint,aps]{revtex}
\tighten
\begin{document}
\preprint{}
\draft
\title{Anyonic states in Chern-Simons theory}
\author{Kurt Haller and Edwin Lim-Lombridas}
\address{Department of Physics, University of Connecticut, Storrs, Connecticut
06269}
\date{\today}
\maketitle
\begin{abstract}
We discuss the canonical quantization of Chern-Simons theory in $2+1$
dimensions, minimally coupled to a Dirac spinor field, first in the temporal
gauge and then in the Coulomb gauge. In our temporal gauge formulation, Gauss's
law and the gauge condition, $A_0 = 0$, are implemented by embedding the
formulation in an appropriate physical subspace. We construct a Fock space of
charged particle states that satisfy Gauss's law, and show that they obey
fermion, not fractional statistics. The gauge-invariant spinor field that
creates these charged states from the vacuum obeys the anticommutation rules
that generally apply to spinor fields. The Hamiltonian, when described in the
representation in which the charged fermions are the propagating particle
excitations that obey Gauss's law, contains an interaction between charge and
transverse current densities. We observe that the implementation of Gauss's law
and the gauge condition does not require us to use fields with graded
commutator algebras or particle excitations with fractional statistics. In our
Coulomb gauge formulation, we implement Gauss's law and the gauge condition,
$\partial_lA_l=0$, by the Dirac-Bergmann procedure. In this formulation, the
constrained gauge fields become functionals of the spinor fields, and are not
independent degrees of freedom. The formulation in the Coulomb gauge confirms
the results we obtained in the temporal gauge: The ``Dirac-Bergmann''
anticommutation rule for the charged spinor fields $\psi$ and $\psi^\dagger$
that have both been constrained to obey Gauss's law, is precisely identical to
the canonical spinor anticommutation rule that generates standard fermion
statistics. And we also show that the Hamiltonians for charged particle states
in our temporal and Coulomb gauge formulations are identical, once Gauss's law
has been implemented in both cases.
\end{abstract}
\narrowtext

\section{Introduction}
In recent work, attention has been focused on anyonic states in gauge theories
with Chern-Simons (CS) interactions
\cite{semenoff,semenoffsodano,luscher,banerjee,matsuyama}. In this work,
gauge-invariant fields have been constructed that create, from a vacuum state,
charged particle states with arbitrary spin and fractional statistics.
Considerable effort has been devoted to understanding the nature of these
states, and the dynamical laws they obey. Some authors have argued that anyons
are a consequence of the imposition of Gauss's law on charged states in $2+1$
dimensional gauge theories with CS terms in their Lagrangian
\cite{semenoff,semenoffsodano,luscher,banerjee,matsuyama}. In these authors'
treatment of this model, local operator-valued fields that create charged
particles from the vacuum obey either purely commutator or anticommutator
algebras; the graded commutator algebras---and the consequent arbitrary spin
and fractional statistics---arise with the construction of nonlocal
gauge-invariant operators, which these authors consider essential for the
imposition of Gauss's law. Other authors have questioned these conclusions
\cite{hagen,hager1989,boyanovsky,ys}. The dynamical implications of the CS term
in gauge theories coupled to charged matter have also been discussed
\cite{matsuyama,jackiwpi,swanson,gwsandlcr}. Jackiw and Pi have shown that CS
fields coupled to charged matter do not generally produce ``pure gauge''
interactions that have no effect on the equations of motion \cite{jackiwpi}.
They point out that, in relativistic quantum field theories, the CS vector
potential cannot be totally gauged away. In nonrelativistic quantum field
theory, in which the CS interaction is pure gauge, Jackiw and Pi have exploited
the pure gauge nature of the CS interaction to remove the entire gauge field
from the Hamiltonian with a gauge transformation. The correspondingly
transformed charged field operator, $\Psi^0({\bf r})$, does not commute with
$\Psi^0({\bf r}^\prime)$, but obeys a graded commutator algebra. $N$-particle
orbitals, represented by appropriately selected matrix elements of products of
these transformed Schr\"odinger field operators, are multivalued. The
constraint imposed by the multivalued boundary condition carries the
information contained in the gauge fields before they were eliminated by the
gauge transformation, and produces charged $N$-particle orbitals that describe
an interacting system of particles. Other investigators have used line
integrals over gauge fields to construct gauge-invariant field operators that
obey graded commutator algebras for relativistic quantum field theories
interacting with a CS field \cite{banerjee}. These authors have identified the
excitations of these gauge-invariant fields as anyonic states with arbitrary
spin and fractional statistics.

In our work, we address this question from a somewhat different point of view.
We investigate a $2+1$ dimensional gauge theory in which the gauge field in
minimally coupled to a charged spinor field. The Lagrangian contains a CS term,
but no Maxwell kinetic-energy term. The gauge field obeys canonical commutation
rules, and the spinor field anticommutation rules. We construct a Fock space of
$N$-particle charged states (the $2+1$ dimensional analogs of electrons and
positrons) that satisfy Gauss's law. In the process, we construct a
gauge-invariant operator-valued spinor field that creates, from a vacuum state,
the charged particle states that satisfy Gauss's law. We demonstrate that this
gauge-invariant spinor fields obeys anticommutation rules; and the excitations
of the gauge-invariant spinor field, which satisfy Gauss's law, obey fermion
rather than fractional statistics. Moreover, it is possible to define these
states so that they change sign in a $2\pi$ rotation, regardless of the value
of the CS coupling constant.

We do not argue that our results invalidate either the anyonic descriptions of
particle states in CS theory, or the gauge invariance of the charged fields
discussed in Ref.~\cite{banerjee}. We do argue that an anyonic description is
not the only possible one for this theory; and in particular, that it is not
required for the implementation of Gauss's law. We demonstrate in this work
that it is possible to formulate a consistent description of the charged
particle excitations as ``normal'' fermions which obey Gauss's law and ordinary
fermion statistics, and which interact through a nonlocal interaction mediated
by the CS field. The availability, in this theory, of a Fock space of states
with normal statistics is consistent with the work of Jackiw and Pi
\cite{jackiwpi}; these authors have shown that nonrelativistic charged bosons
coupled to a CS field can be described by either of two Schr\"odinger field
operators, $\Psi({\bf r})$ or $\Psi^0({\bf r})$. $\Psi({\bf r})$ obeys
``ordinary'' canonical commutation rules, while $\Psi^0({\bf r})$ is subject to
a graded commutator algebra, in which $\Psi^0({\bf r})$ and $\Psi^0({\bf
r}^\prime)$ do not commute. In the representation in which $\Psi({\bf r})$ is
the appropriate Schr\"odinger field operator, explicit nonlocal charged
particle interactions appear in the Hamiltonian. In the representation in which
$\Psi^0({\bf r})$ is the appropriate Schr\"odinger field operator, these
explicit interactions have been replaced by equivalent boundary conditions.
Both representations implement Gauss's law. We observe that in our work, an
explicit interaction between charge and transverse current densities appears in
the Hamiltonian, in the representation in which the electron and positron
operators create (or annihilate) charged particles that obey Gauss's law. A
similar interaction also is reported in Ref.~\cite{jackiwpi} in the
representation in which the orbitals of the boson field $\Psi({\bf r})$ are
used to describe the interacting particles. Our result, that CS theory coupled
to relativistic charged fermions can be formulated in a Fock space of charged
fermion states that satisfy Gauss's law as well as normal statistics, is
consistent with the results of Jackiw and Pi for nonrelativistic charged
bosons.

As has been noted, CS theories do not possess any observable propagating modes
of the gauge field \cite{deser}. Only the charged fermion field gives rise to
observable propagating particle excitations which interact with each other
through their interaction with the gauge field. In our work in the temporal
gauge, we treat this model much as we have previously treated the topologically
massive Maxwell-Chern-Simons (MCS) theory \cite{hl}. We introduce a
gauge-fixing field in such a way that $A_0$ has a conjugate momentum and obeys
canonical commutation rules. Although, as in our treatment of MCS theory,
Gauss's law and the gauge condition are not primary constraints, there are
nevertheless other primary
constraints in CS theory. Primary constraints relate the canonically conjugate
momentum of $A_1$ to $A_2$, and vice versa, so that the constrained gauge field
$A_l$ will be subject to Dirac rather than Poisson commutation rules.
Furthermore, all components of the CS gauge field, $A_1$ and $A_2$ as well as
$A_0$, must be represented entirely in terms of ghost operators, which can
mediate interactions between charges and currents but do not carry
energy-momentum, and have no probability of being observed. Neither
longitudinal nor transverse components of the CS fields have any propagating
particle-like excitations.

In our Coulomb gauge formulation, we implement all constraints, including
Gauss's law and the gauge condition, $\partial_lA_l=0$, by the Dirac-Bergmann
(DB) procedure \cite{dirac,bergmann}. We include a gauge-fixing term,
$-G\partial_lA_l$, in the Lagrangian, in order to provide for the systematic
development of all constraints, including the gauge condition, from the DB
algorithm. In the Coulomb gauge, the gauge fields have no independent degrees
of freedom whatsoever, but are reduced to functionals of the spinor fields. The
constrained fields obey Dirac commutation (anticommutation) rules which must be
evaluated, and which may, and often do, differ from the commutation
(anticommutation) rules of the corresponding unconstrained fields. There is
therefore an opportunity for discrepancies between the commutator
(anticommutator) algebras for constrained and unconstrained fields to arise.
The Dirac anticommutation rules among the constrained spinor fields are of
particular significance, because a graded anticommutation algebra among the
spinor fields may signal the development of ``exotic'' fractional statistics by
their particle excitations.

\section{Formulation of the theory in the temporal gauge}
The Lagrangian for this model is given by
\widetext
\begin{equation}
{\cal L} = \case 1/4 m\epsilon_{ln}(F_{ln}A_0 - 2F_{n0}A_l)
- \partial_0A_0G + j_lA_l - j_0A_0 + \bar{\psi}(i\gamma^\mu
\partial_\mu - M)\psi
\label{eq:L}
\end{equation}
where $F_{ln} = \partial_nA_l - \partial_lA_n$ and $F_{l0} = \partial_lA_0 +
\partial_0A_l$. We follow conventions identical to those in Ref.~\cite{hl}.

The Euler-Lagrange equations are
\begin{equation}
m\epsilon_{ln}F_{n0} - j_l = 0,
\label{eq:ampere1}
\end{equation}
\begin{equation}
\case 1/2 m\epsilon_{ln}F_{ln} + \partial_0G - j_0 = 0,
\label{eq:gaussmotion}
\end{equation}
\begin{equation}
\partial_0A_0 = 0,
\label{eq:DoAo}
\end{equation}
and
\begin{equation}
(M - i\gamma^\mu D_\mu)\psi = 0,
\label{eq:diraceq}
\end{equation}
where $D_\mu$ is the gauge-covariant derivative $D_{\mu} = \partial_{\mu} +
ieA_{\mu}$. Current conservation leads to
\begin{equation}
\partial_0\partial_0G = 0.
\label{eq:DoDoG}
\end{equation}

The momenta conjugate to the fields are given by
\begin{equation}
\Pi_0 = -G,
\label{eq:Pi0}
\end{equation}
and
\begin{equation}
\Pi_l = \case 1/2 m\epsilon_{ln}A_n.
\label{eq:Pii}
\end{equation}
The Hamiltonian density is given by
\begin{equation}
{\cal H} = -\case 1/2 m\epsilon_{ln}F_{ln}A_0 + j_0A_0 -
j_lA_l + {\cal H}_{e\bar{e}}
\label{eq:hdensity}
\end{equation}
where ${\cal H}_{e\bar{e}} = \psi^\dagger(\gamma^0M -
i\gamma^0\gamma^l\partial_l)\psi$ and the total derivative $\partial_n(\case
1/2 m\epsilon_{ln}A_lA_0)$ has been dropped.

The use of the gauge-fixing term $-\partial_0A_0G$ in the Lagrangian ${\cal L}$
leads to the equal-time commutation rule (ETCR)
\begin{equation}
[A_0({\bf x}), G({\bf y})] = -i\delta({\bf x - y}),
\label{eq:AoG}
\end{equation}
and elementary considerations lead to the equal-time anticommutation rule
\begin{equation}
\{\psi_\alpha({\bf x}), \psi_\beta^\dagger({\bf y})\} =
\delta_{\alpha\beta}\delta({\bf x - y}),
\label{eq:psipsidagger}
\end{equation}
for the fermion fields. But the naive use of Eq.~(\ref{eq:Pii}) to set
\begin{equation}
[A_l({\bf x}), \Pi_n({\bf y})] = [A_l({\bf x}), \case 1/2
im\epsilon_{nk}A_k({\bf y})]=i\delta_{ln}\delta({\bf x-y})
\end{equation}
and, after contraction over $\epsilon_{nk}$,
\begin{equation}
[A_l({\bf x}), A_n({\bf y})] =
\frac{2i}{m}\,\epsilon_{ln}\delta({\bf
x - y}),
\end{equation}
is incorrect, because it ignores the fact that $\Pi_l - \case 1/2
m\epsilon_{ln}A_n =0$ constitutes a primary constraint. There are various ways
to arrive at the correct ETCR \cite{dirac,bergmann,faddeev}. One way is to use
the Dirac-Bergmann (DB) procedure \cite{dirac,bergmann}, for which we need the
Poisson commutator matrix
\begin{equation}
{\cal M}_{ln}({\bf x,y}) = [{\cal C}_l({\bf x}), {\cal
C}_n({\bf y})] = -im\epsilon_{ln}\delta({\bf x-y})
\end{equation}
for the primary constraints
\begin{equation}
{\cal C}_l = \Pi_l - \case 1/2 m\epsilon_{ln}A_n.
\label{eq:calCl}
\end{equation}
To implement the DB procedure, we form the total
Hamiltonian density
\begin{equation}
{\cal H}_{\mbox{\rm\scriptsize T}} = {\cal H} + \sum_{l=1}^2 {\cal
C}_l{\cal U}_l
\end{equation}
where the ${\cal U}_l$ are arbitrary $c$-number functions. The commutator
$[H_{\mbox{\rm\scriptsize T}},{\cal C}_i({\bf x})]$ for
$H_{\mbox{\rm\scriptsize T}}=\int d{\bf x}\ {\cal H}_{\mbox{\rm\scriptsize
T}}({\bf x})$, then is
\begin{equation}
[H_{\mbox{\rm\scriptsize T}},{\cal C}_i({\bf x})] = [H,{\cal
C}_i({\bf x})] + \sum_l\int d{\bf y}\ {\cal U}_l({\bf
y})[{\cal C}_l({\bf y}),{\cal C}_i({\bf x})];
\label{eq:HTCi}
\end{equation}
where the brackets represent canonical ``Poisson'' commutators.
Equation~(\ref{eq:HTCi}) leads to
\begin{equation}
m\epsilon_{il}({\cal U}_l + \partial_lA_0)-j_i =0,
\end{equation}
so that ${\cal U}_l$ is identified as ${\cal U}_l = \partial_0A_l$, and no
secondary constraints are generated. Having established that the two primary
constraints given in Eq.~(\ref{eq:calCl}) do not give rise to any secondary
constraints, we recognize the two primary constraints ${\cal C}_l \approx 0$ as
a system of second-class constraints, and use ${\cal Y}({\bf x,y})$, the
inverse of ${\cal M}({\bf x,y})$, to obtain the Dirac commutator
\begin{equation}
[A_l({\bf x}), A_n({\bf y})]^{\mbox{\rm\scriptsize D}} = -
\int d{\bf z}\,d{\bf z}^\prime\ [A_l({\bf x}), C_k({\bf
z})]{\cal Y}_{kk^\prime}({\bf z},{\bf
z}^\prime)[C_{k^\prime}({\bf z}^\prime),A_n({\bf y})].
\end{equation}
The resulting correct expression for the commutator $[A_l({\bf x}), A_n({\bf
y})]$ is the Dirac commutator
\begin{equation}
[A_l({\bf x}), A_n({\bf y})] =
\frac{i}{m}\,\epsilon_{ln}\delta({\bf
x - y}).
\label{eq:AiAj}
\end{equation}
We now construct the following momentum space expansions of the gauge fields in
such a way that the ETCR given in Eqs.~(\ref{eq:AoG}) and (\ref{eq:AiAj}) are
satisfied:
\begin{eqnarray}
A_l({\bf x}) &=& \sum_{\bf k}\frac{k_l}{2m^{3/2}}
\left[a_R({\bf k})e^{i{\bf k \cdot x}} + a_R^\star({\bf
k})e^{-i{\bf k \cdot x}}\right]\nonumber\\
&+& \sum_{\bf
k}\frac{i\sqrt{m}\epsilon_{ln}k_n}{k^2}\left[a_Q({\bf
k})e^{i{\bf k \cdot x}} - a_Q^\star({\bf k})e^{-i{\bf k
\cdot x}}\right]\nonumber\\
&+& \sum_{\bf k}i\phi({\bf k})k_l\left[a_Q({\bf k})e^{i{\bf
k \cdot x}} - a_Q^\star({\bf k})e^{-i{\bf k \cdot
x}}\right],
\label{eq:Ai}
\end{eqnarray}
\begin{equation}
A_0({\bf x}) = \sum_{\bf k}\frac{i}{\sqrt{m}}\left[a_Q({\bf
k})e^{i{\bf k \cdot x}} - a_Q^\star({\bf k})e^{-i{\bf k
\cdot x}}\right],
\label{eq:Ao}
\end{equation}
and
\begin{equation}
G({\bf x}) = -\sum_{\bf k}\frac{\sqrt{m}}{2}\left[a_R({\bf
k})e^{i{\bf k \cdot x}} + a_R^\star({\bf k})e^{-i{\bf k
\cdot x}}\right]
\label{eq:G}
\end{equation}
where $\phi({\bf k})$ is some arbitrary real and even function of ${\bf k}$.
The magnetic field $B$, and the electric field ${\bf E}$ are given by
\begin{equation}
B({\bf x}) =-\sum_{\bf k}\sqrt{m}\left[a_Q({\bf k})e^{i{\bf
k \cdot x}} + a_Q^\star({\bf k})e^{-i{\bf k \cdot
x}}\right].
\label{eq:Bfield}
\end{equation}
and by $E_l = -\partial_lA_0 -i[H,A_l]$ so that
\begin{equation}
E_l({\bf x}) = \sum_{\bf
k}\frac{1}{m}\,\epsilon_{ln}j_n({\bf k})e^{i{\bf k\cdot x}}
\label{eq:Efield}
\end{equation}
as shown in Eq.~(\ref{eq:ampere1}). The explicit form of $\phi({\bf k})$ is
immaterial to the commutation rules given in Eqs.~(\ref{eq:AoG}) and
(\ref{eq:AiAj}); its form as well as its inclusion in Eq.~(\ref{eq:Ai}) are
therefore entirely optional. The operators $a_Q({\bf k})$ and $a_R({\bf k})$
and their Hermitian adjoints $a_Q^\star({\bf k})$ and $a_R^\star({\bf k})$ are
the same ghost operators previously used for the MCS theory \cite{hl}; they
obey the commutation rules
\begin{equation}
[a_Q({\bf k}), a_R^\star({\bf q})] = [a_R({\bf k}),
a_Q^\star({\bf q})] = \delta_{\bf k q},
\label{eq:aQaRstar}
\end{equation}
and
\begin{equation}
[a_Q({\bf k}), a_Q^\star({\bf q})] = [a_R({\bf k}),
a_R^\star({\bf q})] =0.
\label{eq:aQaQstar}
\end{equation}
The use of ghosts is appropriate and necessary for components of gauge fields
which have nonvanishing commutators with each other, but which do not exhibit
any observable, propagating excitations. The representation of the gauge fields
in terms of ghost excitations only, therefore tests the principle that no
observable excitation modes are required to represent the commutation rules
given in Eqs.~(\ref{eq:AoG}) and (\ref{eq:AiAj}).

The Hamiltonian $H =\int d{\bf x}\ {\cal H}({\bf x}) =  H_0
+ H_{\mbox{\scriptsize\rm I}}$, where $H_0$ and
$H_{\mbox{\scriptsize\rm I}}$ are given by
\begin{eqnarray}
H_0 &=& -\int{}d{\bf x}\ \case 1/2 m\epsilon_{ln}F_{ln}A_0 +
H_{e\bar{e}}\nonumber\\
&=& \sum_{\bf k}im\left[a_Q({\bf k})a_Q(-{\bf k}) -
a_Q^\star({\bf k})a_Q^\star(-{\bf k})\right] +
H_{e\bar{e}}
\label{eq:hop}
\end{eqnarray}
with $H_{e\bar{e}} = \int d{\bf x}\ {\cal H}_{e\bar{e}}({\bf
x})$ and
\begin{eqnarray}
H_{\mbox{\scriptsize\rm I}} &=& \sum_{\bf
k}\frac{i}{\sqrt{m}}\left[a_Q({\bf k})j_0(-{\bf k}) -
a_Q^\star({\bf k})j_0({\bf k})\right]\nonumber\\
&-&\sum_{\bf k}\frac{k_l}{2m^{3/2}}\left[a_R({\bf k})j_l(-
{\bf k}) + a_R^\star({\bf k})j_l({\bf
k})\right]\nonumber\\
&-& \sum_{\bf
k}\frac{i\sqrt{m}\epsilon_{ln}k_n}{k^2}\left[a_Q({\bf
k})j_l(-{\bf k}) - a_Q^\star({\bf k})j_l({\bf
k})\right]\nonumber\\
&-& \sum_{\bf k}i\phi({\bf k})k_l\left[a_Q({\bf k})j_l(-{\bf
k}) - a_Q^\star({\bf k})j_l({\bf k})\right].
\label{eq:hip}
\end{eqnarray}
The total Hamiltonian $H_{\mbox{\rm\scriptsize T}}$ reduces to the canonical
Hamiltonian $H$ on the constraint surface on which all ${\cal C}_i$'s are zero,
and it correctly implements time evolution when the Dirac commutation rule
given in
Eq.~(\ref{eq:AiAj}) is used. This can easily be demonstrated by observing that
when the commutators $i[H,A_i]$, $i[H,A_0]$ and $i[H,G]$ are substituted for
$\partial_0A_i$, $\partial_0A_0$ and $\partial_0G$, respectively, the
Euler-Lagrange equations are obtained. The other constraints, $\Pi_0 + G=0$ and
$\Pi_{\text{G}}=0$, have no further effect on the commutation rules for the
gauge fields.

We will implement the gauge constraint, $A_0=0$, and Gauss's law not by using
the DB procedure but, as in earlier work \cite{hl,khd36}, by confining the
dynamical time evolution to an appropriately chosen subspace of the Hilbert
space $\{|h\rangle\}$ in which the Hamiltonian $H$ operates. The Hilbert space
$\{|h\rangle\}$ very closely resembles the Hilbert space used in
Ref.~\cite{hl}; $\{|h\rangle\}$ is based on the perturbative vacuum $|0\rangle$
annihilated by all annihilation operators, $a_Q({\bf k})$ and $a_R({\bf k})$ as
well as the electron and positron annihilation operators $e({\bf k})$ and
$\bar{e}({\bf k})$, respectively. The Hilbert space $\{|h\rangle\}$ contains a
subspace $\{|n\rangle\}$ that consists of all multiparticle electron-positron
states of the form $|N\rangle = \bar{e}^\dagger({\bf q}_1)\cdots
\bar{e}^\dagger({\bf q}_l){e}^\dagger({\bf p}_1)\cdots{e}^\dagger({\bf
p}_n)|0\rangle$, as well as all other states of the form $a_Q^\star({\bf
k}_1)\cdots a_Q^\star({\bf k}_i)|N\rangle$. We note that the commutation rules
for the ghost operators given in Eqs.~(\ref{eq:aQaRstar}) and
(\ref{eq:aQaQstar}) demonstrate that the states $a_Q^\star({\bf k}_1)\cdots
a_Q^\star({\bf k}_i)|N\rangle$ have zero norm, since $\langle N|a_Q({\bf
k}_i)\cdots a_Q({\bf k}_1)a_Q^\star({\bf k}_1)\cdots a_Q^\star({\bf
k}_i)|N\rangle$ can be rewritten as $\langle N|a_Q^\star({\bf k}_i)\cdots
a_Q^\star({\bf k}_1)a_Q({\bf k}_1)\cdots a_Q({\bf k}_i)|N\rangle$ and each of
the $a_Q({\bf k}_i)$ annihilates any state $|N\rangle$. The states in the
subspace $\{|n\rangle\}$ therefore are either free of ghosts, or if they
contain ghosts, they are zero norm states. $H_0$ time translates all states in
$\{|n\rangle\}$ so that they remain contained within it; and the matrix
elements of $H_0$ within $\{|n\rangle\}$, i.e., matrix elements of the form
$\langle n_b|H_0|n_a\rangle$, always vanish when $|n_a\rangle$ or $|n_b\rangle$
contains any $a_Q^\star$ ghosts. States in which $a_R^\star({\bf k})$ operators
act on a state $|n\rangle$, such as $a_R^\star({\bf q}_1)\cdots a_R^\star({\bf
q}_l)a_Q^\star({\bf k}_1)\cdots a_Q^\star({\bf k}_n)|N\rangle$, are included in
$\{|h\rangle\}$, but excluded from $\{|n\rangle\}$. Such states are not
probabilistically interpretable and their appearance in the course of time
evolution signals an inconsistency in the theory. In the next section, we will
show how the implementation of Gauss's law and the gauge choice averts the
development of this inconsistency. Lastly, it should be noted that the unit
operator in the one-particle ghost (OPG) sector is given by
\begin{equation}
1_{\text{OPG}} = \sum_{\bf k}\left[a_Q^\star({\bf
k})|0\rangle\langle 0|a_R({\bf k}) + a_R^\star({\bf
k})|0\rangle\langle 0|a_Q({\bf k})\right].
\label{eq:opg}
\end{equation}
For multiparticle ghost sectors, the obvious generalization of
Eq.~(\ref{eq:opg}) applies.

\section{The Role of Gauss's Law}
As in all other gauge theories, Gauss's law is not an equation of motion in CS
theory. The operator ${\cal G}({\bf x})$ used to implement Gauss's law is
\begin{equation}
{\cal G}({\bf x}) = j_0({\bf x}) -\case 1/2
m\epsilon_{ln}F_{ln}({\bf x}),
\label{eq:gausslawop}
\end{equation}
and whereas $\partial_0 G={\cal G}$, $\partial_0\partial_0 G = \partial_0 {\cal
G} = 0$ is the equation of motion that governs the behavior of this model.
Further measures must be taken to implement ${\cal G}=0$. We can conveniently
express ${\cal G}$ in the form
\begin{equation}
{\cal G}({\bf x}) = \sum_{\bf k}m^{3/2}\left[a_Q({\bf
k})e^{i{\bf k \cdot x}} +
a_Q^\star({\bf k})e^{-i{\bf k \cdot x}}+\frac{j_0({\bf
k})}{m^{3/2}}\,e^{i{\bf k \cdot x}}\right],
\end{equation}
where $j_0({\bf k}) = \int d{\bf x}\ j_0({\bf x})e^{-i{\bf k \cdot x}}$. We can
define an operator $\Omega({\bf k})$ as
\begin{equation}
\Omega({\bf k}) = a_Q({\bf k}) + \frac{1}{2m^{3/2}}j_0({\bf
k}),
\end{equation}
so that
\begin{equation}
{\cal G}({\bf x}) = \sum_{\bf k}m^{3/2}\left[\Omega({\bf
k})e^{i{\bf k \cdot x}} + \Omega^\star({\bf k})e^{-i{\bf k
\cdot x}}\right].
\label{eq:GaussOmega}
\end{equation}
Similarly, we can write $A_0({\bf x})$ as
\begin{equation}
A_0({\bf x}) = \sum_{\bf
k}\frac{i}{\sqrt{m}}\left[\Omega({\bf k})e^{i{\bf k
\cdot x}} - \Omega^\star({\bf k})e^{-i{\bf k \cdot
x}}\right].
\label{eq:AoOmega}
\end{equation}
We can therefore implement Gauss's law and the gauge condition by embedding the
theory in a subspace $\{|\nu\rangle\}$ of another Hilbert space. The subspace
$\{|\nu\rangle\}$ consists of the states $|\nu\rangle$ which satisfy the
condition
\begin{equation}
\Omega({\bf k})|\nu\rangle = 0.
\label{eq:Omeganu}
\end{equation}
It can be easily seen from Eqs.~(\ref{eq:GaussOmega}) and (\ref{eq:AoOmega})
that, in the physical subspace $\{|\nu\rangle\}$, $\langle\nu^\prime|{\cal
G}|\nu\rangle =0$ and $\langle\nu^\prime|A_0|\nu\rangle =0$, so that both
Gauss's law and the gauge condition $A_0 = 0$ hold. Moreover, the condition
$\Omega({\bf k})|\nu\rangle = 0$, once established, continues to hold at all
other times because
\begin{equation}
[H, \Omega({\bf k})] = 0
\label{eq:HOmega}
\end{equation}
so that $\Omega({\bf k})$ is an operator-valued constant. This demonstrates
that a state initially in the physical subspace $\{|\nu\rangle\}$ will always
remain entirely contained within it as it develops under time evolution.

Consider now the unitary transformation $U=e^D$, where
\begin{equation}
D = -i\int{}d{\bf x}\,d{\bf y}\ \sum_{\bf k}\frac{e^{i{\bf k
\cdot (x - y)}}}{k^2}\,\partial_l A_l({\bf x})j_0({\bf y}).
\label{eq:Dconfig}
\end{equation}
It is easy to show that
\begin{equation}
U^{-1}\Omega({\bf k})U = a_Q({\bf k}).
\end{equation}
We can use $U$ to establish a mapping that maps $\Omega({\bf k}) \rightarrow
a_Q({\bf k})$ and $\{|\nu\rangle\} \rightarrow \{|n\rangle\}$, where
$\{|n\rangle\}$ is the subspace described in the preceding section. In this
mapping, operators ${\cal P}$ map into $\tilde{\cal P}$, i.e.,  $U^{-1}{\cal
P}U=\tilde{\cal P}$. For example, $\tilde{\Omega}({\bf k}) = a_Q({\bf k})$, and
$\tilde{H} = U^{-1}HU$ is given by
\begin{eqnarray}
\tilde{H} &=& H_0 - \sum_{\bf
k}\frac{i\epsilon_{ln}k_n}{mk^2}\,j_l({\bf k})j_0(-{\bf
k})\nonumber\\
&-& \sum_{\bf
k}\frac{i\sqrt{m}\epsilon_{ln}k_n}{k^2}\left[a_Q({\bf
k})j_l(-{\bf k}) - a_Q^\star({\bf k})j_l({\bf k})\right].
\label{eq:htilde}
\end{eqnarray}
The similarly transformed fields are
\begin{equation}
\tilde{A}_l({\bf x}) = A_l({\bf x}) - \sum_{\bf k}
\frac{ik_l}{m^{3/2}}\,\phi({\bf k})j_0({\bf k})e^{i{\bf
k\cdot x}} - \sum_{\bf
k}\frac{i\epsilon_{ln}k_n}{mk^2}\,j_0({\bf k})e^{i{\bf
k\cdot x}},
\end{equation}
\begin{equation}
\tilde{A}_0({\bf x}) ={A}_0({\bf x}),\ \ \ \tilde{G}({\bf
x}) = {G}({\bf x})
\end{equation}
and
\begin{equation}
\tilde{\psi}({\bf x}) = \exp\left[{\cal
D}_{\mbox{\rm\scriptsize
U}}({\bf x})\right]\psi({\bf x})
\label{eq:tildepsi}
\end{equation}
where
\begin{equation}
{\cal D}_{\mbox{\rm\scriptsize U}}({\bf x}) = -ie\int d{\bf
y}\ \sum_{\bf k}\frac{e^{i{\bf k \cdot (x-
y)}}}{k^2}\partial_lA_l({\bf y}).
\label{eq:tildepsiD}
\end{equation}
The transformed electric and magnetic fields are
\begin{equation}
\tilde{E}_l({\bf x}) = E_l({\bf x})
\end{equation}
and
\begin{equation}
\tilde{B}({\bf x}) = B({\bf x}) + {\cal B}({\bf x})
\end{equation}
where $E_l({\bf x})$ and $B({\bf x})$ are given by Eqs.~(\ref{eq:Efield}) and
(\ref{eq:Bfield}), respectively,
and
\begin{equation}
{\cal B}({\bf x}) = \frac{j_0({\bf x})}{m}.
\end{equation}
Equations~(\ref{eq:tildepsi}) and (\ref{eq:tildepsiD}) are of particular
importance to one of the questions we are investigating---i.e.,  whether
imposing Gauss's law on the charged particle states of this theory causes them
to develop ``exotic'' fractional statistics. If the anticommutators for the
spinor fields that implement Gauss's law, $\{\tilde{\psi}({\bf x}),
\tilde{\psi}^\dagger({\bf y})\}$ and $\{\tilde{\psi}({\bf x}),
\tilde{\psi}({\bf y})\}$, differ from the canonical spinor anticommutators
$\{{\psi}({\bf x}), {\psi}^\dagger({\bf y})\}=\delta({\bf x}-{\bf y})$ and
$\{{\psi}({\bf x}), {\psi}({\bf y})\}=0$ that account for the fact that the
excitations of $\psi$ and $\psi^\dagger$ are subject to Fermi statistics, then
that difference may signal that the excitations of $\tilde{\psi}$ and
$\tilde{\psi}^\dagger$ are subject to fractional statistics. We also note that
$\tilde{\psi}({\bf x})$ is gauge-invariant; if we gauge transform
$\tilde{\psi}({\bf x})$ within the confines of the temporal gauge, then the
effect of that gauge transformation on ${\cal D}_{\text{U}}({\bf x})$ and on
$\psi({\bf x})$ cancel, so that the spinor field $\tilde{\psi}({\bf x})$ is
gauge-invariant. This gauge invariance is necessary for excitations of
$\tilde{\psi}({\bf x})$ to obey Gauss's law.

To show that the anticommutation rules for $\tilde{\psi}$ and
$\tilde{\psi}^\dagger$ are identical to the anticommutation rules for the
unconstrained ${\psi}$ and ${\psi}^\dagger$, we observe that $\psi({\bf x})$
and $A_l({\bf y})$ (and therefore also $\psi({\bf x})$ and ${\cal
D}_{\text{U}}({\bf y})$) commute at equal times, so that $\{\tilde{\psi}({\bf
x}),\tilde{\psi}^\dagger({\bf y})\}=\delta({\bf x-y})$ and $\{\tilde{\psi}({\bf
x}),\tilde{\psi}({\bf y})\}=0$. The constrained fields $\tilde{\psi}$ and
$\tilde{\psi}^\dagger$ obey the same anticommutation rules as the unconstrained
${\psi}$ and ${\psi}^\dagger$, and are not subject to any exotic graded
anticommutator algebra. The electron and positron  states that implement
Gauss's law therefore obey standard Fermi, not fractional, statistics. This
result can also be demonstrated from the fact the the transformed fields
$\tilde{\psi}$, $\tilde{\psi}^\dagger$, $\tilde{A}_l({\bf x})$ and
$\tilde{\Pi}_l({\bf x})$ are unitarily equivalent to ${\psi}$,
${\psi}^\dagger$, ${A}_l({\bf x})$ and ${\Pi}_l({\bf x})$ respectively, and
that commutators and anticommutators that are equal to $c$-numbers, are
invariant to unitary transformations. It is, of course, important to keep in
mind that the particular form of $\tilde{\psi}({\bf x})$ given in
Eq.~(\ref{eq:tildepsi}) only applies to the temporal gauge and to this method
of quantization. In other gauges, and with other methods of implementing
constraints, the spinor fields that implements Gauss's law will have a
different representation, and questions about the statistics of
electron-positron states that obey Gauss's law arise in a different way. We
will formulate this theory in the Coulomb gauge in later sections of this
paper, and in that work confirm the result that the charged particle states
obey standard Fermi statistics.

It is convenient to establish an entirely equivalent, alternative formalism, in
which all operators and states are unitarily transformed by the unitary
transformation $U$. Since all matrix elements and eigenvalues are invariant to
such a similarity transformation, we can construct a map $\{|\nu\rangle\}
\rightarrow \{|n\rangle\}$, $\Omega({\bf k}) \rightarrow a_Q({\bf k})$, and, in
general, for all other operators ${\cal P}$, ${\cal P} \rightarrow \tilde{\cal
P}$, where $\tilde{\cal P} = U^{-1}{\cal P}U$. We may then use the transformed
representation as an equivalent formulation of the theory, in which Gauss's Law
and the gauge constraint, $A_0=0$, have been implemented. In this equivalent
alternative representation, $\{|n\rangle\}$ is the physical subspace in which
Gauss's law and the gauge condition are implemented, and $\exp(-i\tilde{H}t)$
is the time-translation operator. A time-translation operator will time
translate state vectors entirely within the physical subspace in the
transformed representation, if it is entirely devoid of $a_R^\star({\bf k})$
and $a_R({\bf k})$ operators, or if it contains them at most in the combination
$[a_R^\star({\bf k})a_Q({\bf k}) + a_Q^\star({\bf k})a_R({\bf k})]$. Inspection
of Eq.~(\ref{eq:htilde}) confirms that $\tilde{H}$ is, in fact, entirely devoid
of $a_R^\star({\bf k})$ and $a_R({\bf k})$ operators, so that the
time-translation operator, $\exp(-i\tilde{H}t)$, correctly satisfies this
requirement. Observable multiparticle states in the alternative transformed
representation are described by state vectors in $\{|n\rangle\}$ which we have
previously designated by $|N\rangle$. In CS theory, the only such positive-norm
observable states are charged excitations of the spinor field (we will refer to
them as electrons and positrons for simplicity). The time-translation operator
$e^{-i\tilde{H}t}$ translates state vectors $|N\rangle$ by transforming them
into new state vectors, at a later time $t$; these time-translated state
vectors consist of further positive-norm state vectors $|N^\prime\rangle$, as
well as additional ghost states.  All of the latter are represented by products
of $a_Q^\star({\bf k})$ operators acting on positive-norm states
$|N^\prime\rangle$. At all times, the positive-norm states alone exactly
saturate unitarity. We will refer to a quotient space, which is the set of all
$|N\rangle$, and also is the residue of $\{|n\rangle\}$ after all zero-norm
states have been excised from it.

We can define another Hamiltonian, $\tilde{H}_{\text{quot}}$, which consists of
those parts of $\tilde{H}$ that remain after we have removed all the terms in
which $a_Q^\star({\bf k})$ or $a_Q({\bf k})$ is a factor;
$\tilde{H}_{\text{quot}}$ is given by
\begin{equation}
\tilde{H}_{\text{quot}} = H_{e\bar{e}} - \sum_{\bf
k}\frac{i\epsilon_{ln}k_n}{mk^2}\,j_l({\bf k})j_0(-{\bf
k}).
\end{equation}
The Hamiltonian $\tilde{H}_{\text{quot}}$ contains $H_{e\bar{e}}$, which
describes the kinetic energy of noninteracting electrons and positrons; it also
contains a part that describes a singular nonlocal interaction between the
charge density and the transverse current density. The projections of
$\exp\left[-i\tilde{H}t\right]|N\rangle$ and
$\exp\left[-i\tilde{H}_{\text{quot}}t\right]|N\rangle$ on other state vectors
in the quotient space are identical.  The parts of $\tilde{H}$ that contain
$a_Q^\star({\bf k})$ or $a_Q({\bf k})$ as factors therefore do not play any
role in the time evolution of state vectors within the quotient space of
observable states, and cannot have any effect on the physical predictions of
the theory. The time-evolution operator that time translates physical states in
the quotient space of observable states can therefore be given as
$\exp\left[-i\tilde{H}_{\text{quot}}t\right]$.

If we expand $D$ in momentum space, we get $D = D_1 + D_2$ where
\begin{equation}
D_1 = \sum_{\bf k}\frac{1}{2m^{3/2}}\left[a_R({\bf k})j_0(-
{\bf k}) - a_R^\star({\bf k})j_0({\bf k})\right]
\label{eq:D1}
\end{equation}
and
\begin{equation}
D_2 = \sum_{\bf k}i\phi({\bf k})\left[a_Q({\bf k})j_0(-{\bf
k}) + a_Q^\star({\bf k})j_0({\bf k})\right].
\label{eq:D2}
\end{equation}
Since $D_2$ commutes with $a_Q({\bf k})$, it has no role in transforming
$\Omega({\bf k})$ into $a_Q({\bf k})$, and the operator $V = e^{D_1}$ by itself
achieves the same end as
$U$, i.e.,
\begin{equation}
V^{-1}\Omega({\bf k})V = a_Q({\bf k}).
\end{equation}
We can use $V$ to establish a second mapping of this theory, in which operators
map according to ${\cal P} \rightarrow V^{-1}{\cal P}V = \hat{{\cal P}}$.
$\hat{\Omega}({\bf k}) = a_Q({\bf k})$, so that $\hat{\Omega}$ and
$\tilde{\Omega}$ are identical; under the mapping $P \rightarrow V^{-1}{\cal
P}V = \hat{{\cal P}}$, the subspace $\{|\nu\rangle\}$ maps into the same
subspace $\{|n\rangle\}$ as under the mapping ${\cal P} \rightarrow U^{-1}{\cal
P}U = \tilde{{\cal P}}$. But, in the case of other operators, $\hat{P}$ differs
from $\tilde{{\cal P}}$. For example, $\hat{H}$ is given by
\begin{eqnarray}
\hat{H} &=& H_0 - \sum_{\bf
k}\frac{i\epsilon_{ln}k_n}{mk^2}j_l({\bf k})j_0(-{\bf
k})\nonumber\\
&-& \sum_{\bf k}\frac{i}{m^{3/2}}\,\phi({\bf k})k_lj_l({\bf
k})j_0(-{\bf k})\nonumber\\
&-& \sum_{\bf
k}\frac{i\sqrt{m}\epsilon_{ln}k_n}{k^2}\left[a_Q({\bf
k})j_l(-{\bf k}) - a_Q^\star({\bf k})j_l({\bf
k})\right]\nonumber\\
&-& \sum_{\bf k}i\phi({\bf k})k_l\left[a_Q({\bf k})j_l(-{\bf
k}) - a_Q^\star({\bf k})j_l({\bf k})\right]
\label{eq:hhat}
\end{eqnarray}
Similarly, $\tilde{\psi}$ and $\hat{\psi}$ differ from each other, although
both are gauge-invariant and project, from the correspondingly defined vacuum
states, electron states that implement Gauss's law. $\hat{\psi}$ is given by
$\hat{\psi}({\bf x}) =\exp[{\cal D}_{\mbox{\rm\scriptsize V}}({\bf
x})]\,\psi({\bf x})$, \cite{comm} where
\begin{equation}
{\cal D}_{\mbox{\rm\scriptsize V}}({\bf x}) = -ie\int
d{\bf y}\ \sum_{\bf k}\frac{e^{i{\bf k \cdot (x-
y)}}}{k^2}\left[\partial_lA_l({\bf y}) +
\frac{k^2}{\sqrt{m}}\phi({\bf
k})\epsilon_{ln}\partial_lA_n({\bf y})\right].
\end{equation}
Similarly, $\tilde{J}$ and $\hat{J}$ are the forms into which the Noether
angular momentum operator $J$ is mapped when it is unitarily transformed by $U$
and $V$, respectively. Both these forms, $\tilde{J}$ and $\hat{J}$, are
therefore significant for the rotation of states, and it is of particular
importance to observe that $\tilde{J}$ and $\hat{J}$ differ from each other.
$J$ is given by
\begin{equation}
J = J_{\mbox{\scriptsize\rm g}} + J_{\mbox{\rm\scriptsize
e}},
\label{eq:j}
\end{equation}
where $J_{\mbox{\scriptsize\rm g}}$ and $J_{\mbox{\rm\scriptsize e}}$ are the
angular momenta of the gauge field and the spinors, respectively.
$J_{\mbox{\scriptsize\rm g}}$ and $J_{\mbox{\rm\scriptsize e}}$ are given by
\begin{equation}
J_{\mbox{\scriptsize\rm g}} = -\int d{\bf
x}\,\epsilon_{ln}(\Pi_i x_l\partial_nA_i - G
x_l\partial_nA_0 + \Pi_l A_n)
\end{equation}
and
\begin{equation}
J_{\mbox{\rm\scriptsize e}} = -\int d{\bf x}\ (i\psi^\dagger
x_l\epsilon_{ln}\partial_n\psi+\case 1/2
\psi^\dagger{\gamma_0}\psi).
\label{eq:Je}
\end{equation}
Under the transformation mediated by $U$, $J \rightarrow \tilde{J}$ and
$\tilde{J} = J$, so that $J$ remains untransformed. But, under the
transformation mediated by $V$, $J \rightarrow \hat{J}$ where $\hat{J} = J +
{\cal J}$
and
\begin{eqnarray}
{\cal J} &=& - \sum_{\bf
k}\epsilon_{ln}k_l\,\frac{\partial\phi({\bf k})}{\partial
k_n}\left[a_Q^\star({\bf k})j_0({\bf k}) + a_Q({\bf
k})j_0(-{\bf k})\right]\nonumber\\
&+& \sum_{\bf
k}\frac{\epsilon_{ln}k_l}{2m^{3/2}}\,\frac{\partial\phi({\bf
k})}{\partial k_n}j_0({\bf k})j_0(-{\bf k}).
\end{eqnarray}
We can support the preceding demonstration that $J$ transforms into itself
under the unitary transformation mediated by $U$, whereas it transforms into $J
+ {\cal J}$ under the unitary transformation mediated by $V$, with the
following observation: $D$ is an integral over operators and functions which
all transform as scalars under spatial rotations. Since $J$ is the generator of
spatial rotations, the commutator $[J,D]$ must vanish. $D_1$ is not necessarily
such an integral over scalars, and there is therefore no similar requirement
that $[J,D_1]$ vanishes.

Since $U$ and $V$ map $\Omega({\bf k})$ into $a_Q({\bf k})$ in identical ways,
we can conclude that the implementation of Gauss's law is not responsible for
the fact that $J$ is transformed into $J + {\cal J}$ when $V$ is used to effect
the mapping. In fact, we can use the Baker-Hausdorff-Campbell relation to
construct an operator $W = e^{D^\prime}$, where
\begin{eqnarray}
D^\prime &=& \sum_{\bf k} \frac{i}{2m^{3/2}}\,\phi({\bf
k}){j_0({\bf
k})j_0(-{\bf k})}\nonumber\\
&-& \sum_{\bf k}i\phi({\bf k})\left[a_Q({\bf k})j_0(-{\bf
k}) +
a_Q^\star({\bf k})j_0({\bf k})\right],
\label{eq:Dprime}
\end{eqnarray}
so that $V = UW$. $W$ has the same effect as $V$ on $J$,
i.e., we find that
\begin{equation}
W^{-1}JW = J + {\cal J},
\label{eq:wjw}
\end{equation}
although $W$ leaves $\Omega({\bf k})$ and ${\cal G}({\bf x})$ untransformed and
does not play any role in implementing Gauss's law; $\phi({\bf k})$ is
arbitrary, and if we choose to set $\phi({\bf k}) = 0$, $U$ and $V$ become
identical. If we choose
\begin{equation}
\phi({\bf k}) = \sqrt{m}\,\frac{\delta(k)}{k}\tan^{-
1}\frac{k_2}{k_1},
\label{eq:phik}
\end{equation}
and if we assume that we can carry out the integration over $d{\bf k}$ while
$j_0({\bf k})$ is still operator-valued, then ${\cal J}$ becomes ${\cal J} =
Q^2/4\pi m$, and accounts for the well-known anyonic phase in the rotation of
charged states through $2\pi$.

In comparing $\tilde{H}$ with $\hat{H}$, we note that they differ by some terms
that include $a_Q({\bf k})$ or $a_Q^\star({\bf k})$ as factors. Since both
$\tilde{H}$ and $\hat{H}$ are entirely free of $a_R({\bf k})$ and
$a_R^\star({\bf k})$ operators, $a_Q({\bf k})$ and $a_Q^\star({\bf k})$ commute
with every other operator that appears in $\tilde{H}$ or $\hat{H}$. The terms
which include $a_Q({\bf k})$ or $a_Q^\star({\bf k})$ as factors therefore do
not affect the time evolution of state vectors in the previously defined
quotient space of observable particles (i.e.,  electrons or positrons); they
can neither produce projections on physical states, nor can they contribute
internal loops to radiative corrections. They have no effect whatsoever on the
physical predictions of the theory and if they are arbitrarily amputated from
$\tilde{H}$ or $\hat{H}$, none of the physical predictions of the theory are
affected. The only other difference between $\tilde{H}$ and $\hat{H}$ is
\begin{equation}
h=\sum_{\bf k}\frac{i}{m^{3/2}}\,\phi({\bf k})k_lj_l({\bf
k})j_0(-{\bf k});
\label{eq:smallh}
\end{equation}
$h$ is a total time derivative in $\hat{H}$, which can be
expressed alternatively as $h = i[H_0,\chi]$, as
$h=i[H,\chi]$, as $h=i[\tilde{H},\chi]$, or as
$h=i[\hat{H},\chi]$ where
\begin{equation}
\chi = -\sum_{\bf k}\frac{1}{2m^{3/2}}\,\phi({\bf
k}){j_0({\bf k})j_0(-
{\bf k})}.
\end{equation}
We will discuss the significance of $\chi$ in the next
section.

\section{Rotational Anomalies and Statistics}
Conclusions about the physical implications of this theory depend not only on
the structure of the operators, but also on the properties of the Hilbert space
in which these operators act. The choice of a Hilbert space can have
significant consequences, even though these may not be reflected in the field
equations or the commutation rules. We can, for example, decide to assign the
previously defined Hilbert space $\{|n\rangle\}$ to the representation in which
the Hamiltonian and the angular momentum take the forms $\tilde{H}$, given in
Eq.~(\ref{eq:htilde}), and $\tilde{J}=J$, given in Eq.~(\ref{eq:j})
respectively. We will designate this representation of operators as the
$\tilde{\cal P}$ representation, and the system consisting of operators in the
$\tilde{\cal P}$ representation acting in the Hilbert space $\{|n\rangle\}$ as
the ${\cal U}$ system. This system of operators and states constitutes a theory
in which the charged fermions rotate ``normally,'' acquiring a factor of
$(-1)^N$ in a $2\pi N$ rotation. The sole interaction between electrons in the
${\cal U}$ system is given by
\begin{equation}
{\sf H} =-\sum_{\bf
k}\frac{i\epsilon_{ln}k_n}{mk^2}\,j_l({\bf k})j_0(-{\bf k}).
\label{eq:sfH}
\end{equation}
We can use $W$ to unitarily transform the operators in the $\tilde{\cal P}$
representation so that all operators $\tilde{\cal P}$ are transformed to the
$\hat{\cal P}$ representation, as shown by
\begin{equation}
W^{-1}\tilde{\cal P}W = \hat{\cal P}.
\end{equation}
The Hamiltonian $\hat{H}$, and the angular momentum $\hat{J}$, will then have
the forms given in Eqs.~(\ref{eq:hhat}) and (\ref{eq:wjw}), respectively. By
itself, this change in the form of the operators has no significance. For
example, if we combine these transformed operators with the correspondingly
transformed Hilbert space $\{|n^\prime\rangle\}$, where each
$|n_i^\prime\rangle =W^{-1}|n_i\rangle$, then we have merely regenerated the
${\cal U}$ system of operators and states in another representation. There will
be no rotational anomaly, although ${\cal J}$ appears as part of the angular
momentum $\hat{J}$. The combined transformation of operators and states
guarantees that the rotated state $e^{i(J+{\cal J})\theta}|n_i^\prime\rangle$
returns to $(-1)^N|n_i^\prime\rangle$ in a $2\pi N$ rotation, in spite of the
arbitrary parameter in ${\cal J}$, which appears to imply arbitrary phases in
$2\pi$ rotations. To demonstrate this in detail, we examine
\begin{equation}
\hat{R}(\theta)|n_i^\prime\rangle = e^{i(J+{\cal
J})\theta}|n_i^\prime\rangle = e^{iJ\theta}e^{i{\cal
J}\theta}W^{-1}|n_i\rangle.
\end{equation}
${\cal J}$ commutes with $W^{-1}$; and, for $\phi({\bf k})$
given by Eq.~(\ref{eq:phik}), $J$ commutes with ${\cal J}$,
but not with $W^{-1}$; we can show that
\begin{equation}
e^{iJ\theta}W^{-1} = W^{-1}e^{iJ\theta}e^{-i{\cal J}\theta}
\label{eq:ijtheta}
\end{equation}
so that
\begin{equation}
e^{i(J+{\cal J})\theta}|n_i^\prime\rangle = W^{-
1}e^{iJ\theta}|n_i\rangle
\end{equation}
and the state $|n_i^\prime\rangle$ rotates ``normally,'' to acquire a factor of
$(-1)^N$ in a $2\pi N$ rotation, as is required by the similarity
transformation.

Alternatively, we could just as well assign the Hilbert space $\{|n\rangle\}$
to the $\hat{\cal P}$ representation of operators. We will designate the system
consisting of operators in the $\hat{\cal P}$ representation and the Hilbert
space $\{|n\rangle\}$ as the ${\cal W}$ system. The systems ${\cal W}$ and
${\cal U}$ both implement the equations of motion, as well as Gauss's law and
the gauge choice $A_0=0$. There is no reason to prefer one system over the
other on the basis of dynamics or constraints. In the ${\cal W}$ system of
operator and states, however, the ${\cal J}$ in $\hat{J}$ would be responsible
for an anomalous rotational phase that we associate with anyonic behavior.
Subsequent similarity transformations that transform to the operator
representation $\tilde{\cal P}$ and the states $W|n_i\rangle$, would preserve
that rotational phase anomaly, although then the arbitrary parameter would not
reside in $J$. We can conclude from these observations that rotational phase
anomalies are possible; but they are not an inevitable feature of this gauge
theory. However, contrary to what has been suggested by some authors
\cite{semenoff,semenoffsodano,gwsandlcr}, it is not the implementation of
Gauss's law that is responsible for anyonic rotational phases. Gauss's law can
be implemented with or without producing arbitrary rotational phases. Moreover,
in corroboration of a result obtained by other means \cite{hagen}, we find that
regardless of whether the arbitrary rotational phase develops, the
anticommutation rule that governs the electron field operator remains unchanged
by the unitary transformations ($U$ or $V$) that are instrumental in
implementing Gauss's law in the $\{|n\rangle\}$ space. And that observation
applies equally to the free Dirac field and to the gauge-invariant electron
field that projects electrons that obey Gauss's law.  The ``normal'' and the
``anyonic'' operators are unitarily equivalent and both obey Fermi-Dirac
statistics. Graded commutator algebras and ``exotic'' fractional statistics do
not arise in the process of implementing Gauss's law and establishing the
${\cal U}$ and ${\cal W}$ systems of operators and states.

We next turn our attention to the extra term, $h$, that appears in the
Hamiltonian $\hat{H}$ in the ${\cal W}$ system; $h$ is the only part of
$(\hat{H}-\tilde{H})$ that describes interactions between charges and currents,
and which therefore might possibly account for physically observable
discrepancies between $\tilde{H}$ and $\hat{H}$. Since the ${\cal U}$ and
${\cal W}$ systems of operators and states both implement the same dynamical
equations and constraints, the question whether both these systems make
identical physical predictions is of considerable interest. The fact that $h$
given in Eq.~(\ref{eq:smallh}) is a total time-derivative gives us {\em a
priori} confidence that it will not affect the $S$-matrix produced by this
theory. A formal argument that confirms this result is based on a theorem about
the relationship between two representations, in which the Hamiltonian in one
is a unitary transform of the Hamiltonian in the other, but the states in both
representations are left untransformed, and are identical \cite{khd36,el3}.
When this theorem is applied to the question we have raised here, we find that
the on-shell transition amplitudes determined by the two Hamiltonians are
related by
\begin{equation}
\tilde{T}_{f,i} = \hat{T}_{f,i} + i\epsilon R
\label{eq:righthand}
\end{equation}
where $\tilde{T}_{f,i}$ and $\hat{T}_{f,i}$ are the scattering amplitudes
\begin{equation}
\tilde{T}_{f,i} = \langle
f|\left(\tilde{H}_{\mbox{\rm\scriptsize I}} +
\tilde{H}_{\mbox{\rm\scriptsize I}}\frac{1}{E-
\tilde{H}+i\epsilon}\tilde{H}_{\mbox{\rm\scriptsize
I}}\right)|i\rangle
\end{equation}
and
\begin{equation}
\hat{T}_{f,i} = \langle
f|\left(\hat{H}_{\mbox{\rm\scriptsize I}} +
\hat{H}_{\mbox{\rm\scriptsize I}}\frac{1}{E-
\hat{H}+i\epsilon}\hat{H}_{\mbox{\rm\scriptsize
I}}\right)|i\rangle;
\end{equation}
$\tilde{H}_{\mbox{\rm\scriptsize I}}$ and
$\hat{H}_{\mbox{\rm\scriptsize I}}$ are given by
\begin{equation}
\tilde{H}_{\mbox{\rm\scriptsize I}} = \tilde{H} - H_0
\end{equation}
and
\begin{equation}
\hat{H}_{\mbox{\rm\scriptsize I}} = \hat{H} - H_0;
\end{equation}
and $R$ is given by
\begin{equation}
R = \langle f|\left(\tilde{H}_{\mbox{\rm\scriptsize
I}}\frac{1}{E-\tilde{H}+i\epsilon}(1-e^{-i\chi}) - (1-e^{-
i\chi})\frac{1}{E-
\hat{H}+i\epsilon}\hat{H}_{\mbox{\rm\scriptsize
I}}\right)|i\rangle;
\end{equation}
$|i\rangle$ and $|f\rangle$ are two multiparticle electron-positron states in
$\{|n\rangle\}$ with identical energy $E$, and represent initial and final
states in scattering events respectively. We note that if $H_0$ and $\chi$
commute, then $h$ vanishes. If $H_0$ and $\chi$ fail to commute, then $R$ will
consist of terms proportional to
\begin{displaymath}
\int dE_n\ \langle f|\tilde{H}_{\mbox{\rm\scriptsize
I}}+\tilde{H}_{\mbox{\rm\scriptsize I}}(E-
\tilde{H}+i\epsilon)^{-1}\tilde{H}_{\mbox{\rm\scriptsize
I}}|n\rangle\langle n|(-i\chi)^k|i\rangle(E-
E_n+i\epsilon)^{-1}
\end{displaymath}
and
\begin{displaymath}
\int dE_n\ \langle f|(-i\chi)^k|n\rangle\langle
n|\hat{H}_{\mbox{\rm\scriptsize
I}}+\hat{H}_{\mbox{\rm\scriptsize I}}(E-
\hat{H}+i\epsilon)^{-1}\hat{H}_{\mbox{\rm\scriptsize
I}}|i\rangle(E-E_n+i\epsilon)^{-1}
\end{displaymath}
where $k$ is an integer. When $\chi$ and $H_0$ do not commute, $R$ will not
give rise to any $1/i\epsilon$ singularities to cancel the $i\epsilon$ on the
right hand side of Eq.~(\ref{eq:righthand}), except perhaps, for contributions
that represent self-energy insertions in external lines. These contributions
only affect renormalization constants and do not affect physical quantities.

We conclude, therefore, that the physical predictions of this model, i.e., the
$S$-matrix elements that determine scattering amplitudes and energy level
shifts for electron-positron systems, are insensitive to whether the charged
states develop anomalous phases.

\section{Formulation of the theory in the Coulomb gauge}
To confirm the results we obtained in the preceding discussion of CS theory in
the temporal gauge, we now also formulate the same model in the Coulomb gauge.
The Coulomb gauge formulation makes use of a quantization procedure that is
different from the one we used for the temporal gauge, and therefore can
provide independent corroboration of our earlier conclusions. In the Coulomb
gauge, the gauge field $A_0$ is not involved in the gauge condition, so that a
gauge-fixing term cannot be used to generate a canonical momentum conjugate to
$A_0$. The most convenient ways to quantize CS theory in the Coulomb gauge are
the Dirac-Bergmann (DB) procedure \cite{dirac,bergmann} and the symplectic
method of Faddeev and Jackiw \cite{faddeev}. We will here use the DB procedure
to impose the constraints. In this method, the canonical ``Poisson''
commutators (anticommutators) are replaced by their respective Dirac
commutators (anticommutators), which apply to the fields that obey all the
constraints of the theory. Since the Dirac and the canonical commutators
(anticommutators) can, and often do, differ from each other, this method
enables us to investigate whether the Dirac anticommutator for the spinor field
$\psi$ and its adjoint $\psi^\dagger$ differ from the corresponding canonical
anticommutator. A discrepancy between the Dirac and canonical anticommutators
for the spinor fields could signal the development of ``exotic'' fractional
statistics due to the imposition of Gauss's law. On the other hand, identity of
the Dirac and the canonical anticommutators for the spinor fields demonstrate
that the excitations of the charged spinor field that obey Gauss's law (as well
as all other constraints) also obey standard Fermi statistics. The question,
whether the imposition of Gauss's law produces charged particle excitations
that are subject to exotic statistics, therefore arises in a new way in the
Coulomb gauge. In this section, we will carry out this quantization procedure
and demonstrate explicitly that the implementation of Gauss's law for the
charged spinor field does not change the anticommutation rule for $\psi$ and
$\psi^\dagger$, and does not cause the excitations of these fields to develop
exotic fractional statistics.

The Lagrangian for CS theory in the Coulomb gauge is given by
\begin{equation}
{\cal L} = \case 1/4 m\epsilon_{ln}(F_{ln}A_0 - 2F_{n0}A_l) - G\partial_lA_l +
j_lA_l - j_0A_0 + \bar{\psi}(i\gamma^\mu\partial_\mu - M)\psi.
\label{eq:LD}
\end{equation}
This Lagrangian differs from Eq.~(\ref{eq:L}) only in that the gauge-fixing
term $-G\partial_0A_0$ is replaced by $-G\partial_lA_l$. We have included a
gauge-fixing term for the Coulomb gauge in Eq.~(\ref{eq:LD}) to enable us to
develop all the constraints systematically from the Lagrangian.

The Euler-Lagrange equations generated by the Lagrangian are
\begin{equation}
m\epsilon_{ln}F_{n0} - j_l - \partial_l G= 0,
\label{eq:ampere1C}
\end{equation}
\begin{equation}
\case 1/2 m\epsilon_{ln}F_{ln} - j_0 = 0,
\label{eq:gaussmotionC}
\end{equation}
\begin{equation}
\partial_lA_l = 0,
\label{eq:DlAlC}
\end{equation}
and
\begin{equation}
(M - i\gamma^\mu D_\mu)\psi = 0.
\label{eq:diraceqC}
\end{equation}
The momenta conjugate to the gauge fields are $\Pi_l = \case 1/2
m\epsilon_{ln}A_n$, for $l=1,2$; $\Pi_0=0$, where $\Pi_0$ is the momentum
conjugate to $A_0$; and $\Pi_{\text{G}}=0$, where $\Pi_{\text{G}}$ is the
momentum conjugate to the gauge-fixing field $G$. For the spinor fields, we
obtain the momenta $\Pi_\psi = i\psi^\dagger$ and $\Pi_{\psi^\dagger}=0$ which
are conjugate to $\psi$ and $\psi^\dagger$, respectively. The corresponding
primary constraints can be expressed as ${\cal C}_i\approx 0$ for
$i=1,\ldots,4$, as well as ${\cal C}_\psi \approx 0$ and ${\cal
C}_{\psi^\dagger}\approx 0$, where
\begin{equation}
{\cal C}_1 = \Pi_1 - \case 1/2 m A_2,
\label{eq:C1}
\end{equation}
\begin{equation}
{\cal C}_2 = \Pi_2 + \case 1/2 m A_1,
\label{eq:C2}
\end{equation}
\begin{equation}
{\cal C}_3 = \Pi_0,
\label{eq:C3}
\end{equation}
\begin{equation}
{\cal C}_4 = \Pi_{\text{G}},
\label{eq:C4}
\end{equation}
\begin{equation}
{\cal C}_\psi = \Pi_\psi - i\psi^\dagger,
\label{eq:Cpsi}
\end{equation}
and
\begin{equation}
{\cal C}_{\psi^\dagger} = \Pi_{\psi^\dagger}.
\label{eq:Cpsidagger}
\end{equation}

The total Coulomb gauge Hamiltonian, $H_{\text{T}}{}^{\text{C}}$, is given by
\begin{eqnarray}
H_{\text{T}}{}^{\text{C}} &=& \int d{\bf x}\ \psi^\dagger(\gamma_0M -
i\gamma_0\gamma_l\partial_l)\psi + \int d{\bf x}\ \left[-\case 1/2
m\epsilon_{ln}F_{ln}A_0 + G\partial_lA_l\right.\nonumber\\
&+& \left.j_0A_0 - j_lA_l - {\cal U}_1{\cal C}_1-{\cal U}_2{\cal C}_2 - {\cal
U}_3{\cal C}_3 - {\cal U}_4{\cal C}_4- {\cal U}_\psi{\cal C}_\psi - {\cal
U}_{\psi^\dagger}{\cal C}_{\psi^\dagger}\right]
\label{eq:HextC}
\end{eqnarray}
where ${\cal U}_1, \ldots, {\cal U}_4$ designate arbitrary functions that
commute with all operators; ${\cal U}_\psi$ and ${\cal U}_{\psi^\dagger}$
designate arbitrary functions that are Grassmann numbers, which anticommute
with all fermion fields and with Grassmann numbers, but commute with bosonic
operators and with ${\cal U}_1, \ldots, {\cal U}_4$. In the imposition of
constraints, we will use the Poisson bracket, $[\![A,B]\!]$, of two operators
$A$ and $B$, defined as $[\![A,B]\!] = AB - (-1)^{n(A) n(B)}BA$, where $n(P)$
is an index for the operators $P$;\footnote{We generally follow the conventions
in Sundermeyer \cite{sundermeyer}. The definition of Poisson bracket used here,
however, differs from Sundermeyer's definition by a factor $i$.} $n(P)=0$ if
$P$ is a bosonic operator, such as a gauge field or a bilinear combination of
fermion fields; and $n(P)=1$ if $P$ is a Grassmann number, or a fermionic
operator such as $\psi$ or $\psi^\dagger$. The Poisson bracket $[\![A,B]\!]$ is
the commutator $[A,B]$ when $A$ and $B$ are both bosonic operators, or if one
is bosonic and the other fermionic. But $[\![A,B]\!]$ is the anticommutator
$\{A,B\}$ when $A$ and $B$ are both fermionic operators.

We use the total Hamiltonian to generate the further constraints needed to
maintain the stability of the primary constraints under time evolution. For
this purpose, we evaluate time derivatives of the primary constraints by using
the equation $\partial_0{\cal C}_i = [\![H_{\text{T}}{}^{\text{C}},C_i]\!]$,
and set $\partial_0C_i \approx 0$. In this way, we find that $\partial_0C_1
\approx 0$ and $\partial_0C_2 \approx 0$ lead to equations for ${\cal U}_1$ and
${\cal U}_2$, but do not generate any secondary constraints. The equation
$\partial_0{\cal C}_3 \approx 0$ leads to the secondary constraint ${\cal C}_5
\approx 0$ where
\begin{equation}
{\cal C}_5 = m\epsilon_{ln}\partial_lA_n + j_0,
\label{eq:C5}
\end{equation}
which implements Gauss's law. $\partial_0{\cal C}_5 \approx 0$ leads to a
further, tertiary constraint, ${\cal C}_6 \approx 0$ where
\begin{equation}
{\cal C}_6 = \partial_l\partial_l G.
\end{equation}
This constraint is necessary for consistency between Eq.~(\ref{eq:ampere1C})
and Gauss's law. To demonstrate this, we observe that taking the
two-dimensional divergence of Eq.~(\ref{eq:ampere1C}), and applying current
conservation, lead to $\partial_0\left[\case 1/2
m\epsilon_{ln}F_{ln}-j_0\right] + \partial_l\partial_lG =0$, which is
inconsistent with Gauss's law unless $\partial_l\partial_lG=0$. The tertiary
constraint ${\cal C}_6=\partial_l\partial_lG\approx 0$ ends this particular
chain of unfolding constraints; imposing the condition that
$\partial_0(\partial_l\partial_lG)\approx 0$, leads to an equation for ${\cal
U}_4$, but generates no further constraints. To insure the stability of the
constraint $\Pi_{\text{G}}\approx 0$, we set $\partial_0\Pi_{\text{G}}\approx
0$ and obtain ${\cal C}_7\approx 0$ where
\begin{equation}
{\cal C}_7 = \partial_lA_l,
\end{equation}
which implements the gauge condition for the Coulomb gauge. The equation
$\partial_0{\cal C}_7 \approx 0$ leads to the tertiary constraint ${\cal C}_8
\approx 0$ where
\begin{equation}
{\cal C}_8 = \partial_l\partial_lA_0 + \frac{1}{m}\,\epsilon_{ln}\partial_lj_n.
\end{equation}
The constraint ${\cal C}_8 \approx 0$ is an expression of the fact that
Eq.~(\ref{eq:ampere1C}), which is an equation of motion for the gauge field
$A_l$ in most gauges, reduces to a constraint in the Coulomb gauge.
Equation~(\ref{eq:ampere1C}), which has a longitudinal as well as a transverse
component, can be expressed as $m\epsilon_{ln}(\partial_nA_0 +
\partial_0A_n)-j_l - \partial_lG=0$. The longitudinal component has just been
shown to lead to the constraint $\partial_l\partial_lG \approx 0$. The
transverse component can be extracted by contracting over
$\epsilon_{il}\partial_i$, and noting that
$\epsilon_{il}\epsilon_{ln}=-\delta_{in}$. In the resulting equation, the sole
remaining time derivative, $\partial_0\partial_lA_l$, vanishes because of the
Coulomb gauge condition, leaving ${\cal C}_8\approx 0$ as a constraint. The
equation $\partial_0{\cal C}_8\approx 0$ leads to no further constraint, so
that ${\cal C}_8\approx 0$ terminates the chain of constraints that develops
from $\Pi_{\text{G}}\approx 0$. In the case of the spinor fields, the equations
$\partial_0C_\psi\approx 0$ and $\partial_0C_{\psi^\dagger}\approx 0$ lead to
equations for the Grassmann functions ${\cal U}_\psi$ and ${\cal
U}_{\psi^\dagger}$; but they do not lead to any further constraints.

The preceding analysis leads to ten second-class constraints for this gauge
theory. Imposition of the constraints requires that we form the matrix ${\cal
M}({\bf x},{\bf y})$, whose elements are ${\cal M}_{ij}({\bf x},{\bf
y})=[\![{\cal C}_i({\bf x}),{\cal C}_j({\bf y})]\!]$. We assign the values
${\cal C}_1, \ldots, {\cal C}_{10}$ to the descending horizontal rows of the
matrix, as well as to the sequence of vertical columns, where ${\cal C}_1,
\ldots, {\cal C}_8$ refer to the previously defined constraints; for simplicity
we will designate ${\cal C}_\psi$ and ${\cal C}_{\psi^\dagger}$ as ${\cal C}_9$
and ${\cal C}_{10}$, respectively. The matrix ${\cal M}({\bf x},{\bf y})$ is
evaluated and inverted; its inverse, ${\cal Y}({\bf x},{\bf y})$, which obeys
\begin{equation}
\int d{\bf z}\ {\cal M}_{ik}({\bf x},{\bf z}){\cal Y}_{kj}({\bf z},{\bf
y})=\int d{\bf z}\ {\cal Y}_{ik}({\bf x},{\bf z}){\cal M}_{kj}({\bf z},{\bf y})
= \delta_{ij}\delta({\bf x}-{\bf y})
\end{equation}
is used to calculate the Dirac commutators (anticommutators) by applying the
equation
\begin{equation}
[\![\xi({\bf x}),\zeta({\bf y})]\!]^{\text{D}} = [\![\xi({\bf x}),\zeta({\bf
y})]\!] - \sum_{i,j=1}^{10}\int d{\bf z}\,d{\bf z}^\prime\ [\![\xi({\bf
x}),{\cal C}_i({\bf z})]\!]{\cal Y}_{ij}({\bf z},{\bf z}^\prime)[\![{\cal
C}_j({\bf z}^\prime),\zeta({\bf y})]\!].
\end{equation}
We observe that the resulting Dirac commutators (anticommutators) are given by
\begin{equation}
[\![\psi({\bf x}),\psi^\dagger({\bf y})]\!]^{\text{D}} = \{\psi({\bf
x}),\psi^\dagger({\bf y})\}=\delta({\bf x}-{\bf y}),
\label{eq:psipsidaggerD}
\end{equation}
\begin{equation}
[\![\psi({\bf x}),\psi({\bf y})]\!]^{\text{D}} = \{\psi({\bf x}),\psi({\bf
y})\}=0,
\label{eq:psipsiD}
\end{equation}
\begin{equation}
[\![A_0({\bf x}),\psi({\bf y})]\!]^{\text{D}} = [A_0({\bf x}),\psi({\bf y})] =
-\frac{ie}{m}\,\epsilon_{ln}\frac{(x-y)_l}{2\pi|{\bf
x-y}|^2}\,\gamma_0\gamma_n\psi({\bf y}),
\label{eq:aopsicom}
\end{equation}
\begin{equation}
[\![A_l({\bf x}),\psi({\bf y})]\!]^{\text{D}} = [A_l({\bf x}),\psi({\bf y})] =
-\frac{ie}{m}\,\epsilon_{ln}\frac{(x-y)_n}{2\pi|{\bf x-y}|^2}\,\psi({\bf y}),
\end{equation}
\begin{equation}
[\![A_l({\bf x}),A_0({\bf y})]\!]^{\text{D}}= [A_l({\bf x}),A_0({\bf y})]=0
\end{equation}
and
\begin{equation}
[\![A_l({\bf x}),A_n({\bf y})]\!]^{\text{D}} = [A_l({\bf x}),A_n({\bf y})] = 0.
\label{eq:alancom}
\end{equation}
Equations~(\ref{eq:psipsidaggerD}) and (\ref{eq:psipsiD}) demonstrate that the
constrained spinor field obeys standard anticommutation rules, and not a graded
anticommutator algebra; and that the charged excitations of that spinor field
are subject to standard Fermi statistics, and not the exotic fractional
statistics that would result from a graded anticommutator algebra. In contrast
to the spinor field, the Dirac commutators of the gauge fields differ
substantially both from the unconstrained canonical commutators, and also from
their corresponding values in the temporal gauge. The observation that the
spinor anticommutation rule is unaffected by constraints, and identical in the
Coulomb and temporal gauges, therefore is not trivial.

The Dirac commutators (anticommutators) imply relationships among the
constrained operators that reduce the independent degrees of freedom of this
theory. There are various procedures for making these relationships explicit,
but in this model the simplest way is to use the constraint equations to
express the gauge fields as functionals of the charge and current densities.
Since the gauge fields have no observable, propagating degrees of freedom, this
procedure can completely eliminate all gauge fields from the Hamiltonian. We
find, for example, that ${\cal C}_5 = 0$ can be solved to yield
\begin{equation}
A_l({\bf x}) =
-\frac{1}{m}\,\epsilon_{ln}\frac{1}{\nabla^2}\,\frac{\partial}{\partial
x_n}\,j_0({\bf x}),
\label{eq:AlD}
\end{equation}
and ${\cal C}_8=0$ can be solved for
\begin{equation}
A_0 = -\frac{1}{m}\,\epsilon_{ln}\frac{1}{\nabla^2}\,\partial_lj_n.
\label{eq:A0D}
\end{equation}
When we use Eqs.~(\ref{eq:AlD}) and (\ref{eq:A0D}) and the Fermion
anticommutation relations for the constrained $\psi$ and $\psi^\dagger$, we
exactly reproduce the Dirac commutators given in
Eqs.~(\ref{eq:aopsicom})--(\ref{eq:alancom}). The value $G^{\text{D}}$ of the
gauge-fixing field $G$ on the constraint surface can be shown to be zero using
the relation
\begin{equation}
G^{\text{D}}({\bf x}) = G({\bf x}) - \sum_{i,j=1}^{10}\int d{\bf y}\,d{\bf z}\
[\![G({\bf x}),{\cal C}_i({\bf y})]\!]{\cal Y}_{ij}({\bf y},{\bf z}){\cal
C}_j({\bf z}).
\end{equation}
Since $G^{\text{D}}=0$, then Eq.~(\ref{eq:ampere1C}) reduces to
$m\epsilon_{ln}F_{0n}-j_l=0$ under the influence of the constraints. When these
constrained representations of the gauge fields are substituted into the
Hamiltonian, and all the other constraint functions, ${\cal C}_i$, for
$i=1,\ldots,10$, are set to zero, we obtain the result that
\begin{equation}
H_{\text{C}} = \int d{\bf x}\ \psi^\dagger(\gamma_0M -
i\gamma_0\gamma_l\partial_l)\psi - \sum_{\bf
k}\frac{i\epsilon_{ln}k_n}{mk^2}\,j_l({\bf k})j_0(-{\bf k}).
\end{equation}
We observe that this form of the Hamiltonian is exactly identical to
$\tilde{H}_{\text{quot}}$ in the temporal gauge. This exact identity of
$H_{\text{C}}$ and $\tilde{H}_{\text{quot}}$ provides a very compelling
demonstration that the charged states that obey Gauss's law are subject to the
identical dynamics in both our temporal gauge and Coulomb gauge formulations of
this model. Since the gauge-independence of the physical predictions of this
theory is a firm requirement, this identity serves as a significant
corroboration of the consistency and correctness of both of our formulations of
this model. Similarly, our separate demonstrations, in the temporal gauge and
the Coulomb gauge formulations, of the anticommutation rules for the spinor
fields that create and annihilate charged particles that obey Gauss's law, show
that these charged particle states obey standard Fermi rather than fractional
statistics. The formalism employed for each of these demonstrations is specific
to each gauge---we observe, for example, that ${\cal D}_{\text{U}}({\bf x})$,
given in Eq.~(\ref{eq:tildepsiD}), whose properties play an essential role in
the argument that pertains to the temporal gauge, would vanish in the Coulomb
gauge. But the fact that the same result is reached in both gauges, confirms
that our conclusion about the statistics of charged particle states is
gauge-independent, and makes the argument particularly persuasive.

\section{Discussion}
In the preceding work, we have demonstrated that in Chern-Simons theory coupled
to a charged-fermion field, the imposition of Gauss's law does not cause the
spinor fields to develop an ``exotic'' graded commutator algebra. The charged
excitations of these fields are subject to standard Fermi, and not exotic
fractional statistics, even when they obey Gauss's law. In order to provide
convincing arguments for these conclusions, we have carried out detailed
calculations that support them, both in the temporal and in the Coulomb gauge
formulations of the theory. We have also obtained a time evolution operator for
the single and multiparticle electron-positron states that obey Gauss's law,
and have shown that this time evolution operator is identical in the two
calculations we have carried out, one in the temporal and the other in the
Coulomb gauge. This result provides further confirmation that the quantization
procedures and the conclusions based on them are correct. We have also shown
that the charged states may or may not acquire arbitrary phases in $2\pi$
rotations, depending upon the way we choose to represent them. However, that
choice of representation can be made independently of whether the charged
states of the theory obey Gauss's law; it is not a consequence of the
imposition of the constraints. Moreover, the choice of representation that
determines whether arbitrary rotational phases result from $2\pi$ rotations
does not have any implications for the physical predictions of the theory.

\acknowledgements
This research was supported by the Department of Energy
under Grant No. DE-FG02-92ER40716.00.


\begin{references}
\bibitem{semenoff}G. W. Semenoff, in {\it Physics, Geometry, and Topology\/},
H. C. Lee, Ed., (Plenum Press, New York, 1990); G. W. Semenoff, Phys. Rev.
Lett. {\bf 61}, 517 (1988).
\bibitem{semenoffsodano}G. W. Semenoff and P. Sodano, Nucl. Phys. {\bf B328},
753 (1989).
\bibitem{luscher}M. L\"uscher, Nucl. Phys. {\bf B326}, 557 (1989).
\bibitem{banerjee}R.~Banerjee, Phys. Rev. Lett. {\bf 69}, 17 (1992); Phys. Rev.
D {\bf 48} 2905 (1993); A.~Foerster and H.~O.~Girotti, Phys. Lett. {\bf B 230},
83 (1989).
\bibitem{matsuyama}T.~Matsuyama, Phys.~Rev.~D {\bf 42}, 3469 (1990); {\bf 44},
2616 (1991).
\bibitem{hagen}C. R. Hagen, Phys. Rev. D {\bf 31}, 2135 (1985).
\bibitem{hager1989}C.~R.~Hagen, Phys.~Rev.~Lett.~{\bf 63}, 1025 (1990).
\bibitem{boyanovsky}D. Boyanovsky, Phys. Rev. D {\bf 42}, 1179 (1990); D.
Boyanovsky, E. T. Newman, and C. Rovelli, {\it ibid.\/} {\bf 45}, 1210 (1992).
\bibitem{ys}Y. Srivastava and A. Widom, Phys. Rep. {\bf 148}, 1 (1987).
\bibitem{jackiwpi}R.~Jackiw and So-Young Pi, Phys. Rev. D {\bf 42}, 3500
(1990).
\bibitem{swanson}M. Swanson, Phys. Rev. D {\bf 42}, 552 (1990).
\bibitem{gwsandlcr}G. W. Semenoff and L. C. R. Wijewardhana, Phys. Lett. B {\bf
184}, 397 (1987).
\bibitem{deser}C. Hagen, Ann. Phys. (N.Y.) {\bf 157}, 342 (1984).
\bibitem{hl}K. Haller and E. Lim-Lombridas, Phys. Rev. D {\bf 46}, 1737 (1992).
\bibitem{dirac}P. A. M. Dirac, {\it Lectures on Quantum Mechanics\/} (Yeshiva
University Press, New York, 1964).
\bibitem{bergmann}P.~G.~Bergmann and I.~Goldberg, Phys. Rev. {\bf 98}, 531
(1955).
\bibitem{faddeev}L. D. Faddeev and R. Jackiw, Phys. Rev. Lett. {\bf 60}, 1692
(1988); G. V. Dunne, R. Jackiw, and C. A. Trugenberger, Ann. Phys. (N.Y.) {\bf
194}, 197 (1989).
\bibitem{khd36} K. Haller, Phys. Rev. D {\bf 36}, 1830 (1987).
\bibitem{comm}We note that the gauge-invariant operators used by Boyanovsky
{\em et al.} in Ref.~\cite{boyanovsky} to project charged particle states that
implement Gauss's law are similar to our $\tilde{\psi}$ (but not to our
$\hat{\psi}$). Their results are in agreement with our observation that the
implementation of Gauss's law does not affect the anticommutation rules for
charged fermions.
\bibitem{el3}K.~Haller and E.~Lim-Lombridas, Found. of Phys. {\bf 24}, 217
(1994).
\bibitem{sundermeyer}K.~Sundermeyer, {\it Constrained Dynamics\/} (Springer,
New York, 1982).
\end{references}
\end{document}